

\def\singlespace{\normalbaselines}
\def\oneandahalfspace{\baselineskip=1.15\normalbaselineskip plus 1pt
\lineskip=2pt\lineskiplimit=1pt}

\def\np{\vfill\eject}
\def\nl{\hfil\break}

\def\nofirstpagenoten{\nopagenumbers\footline={\ifnum\pageno>1\tenrm
\hss\folio\hss\fi}}
\def\nofirstpagenotwelve{\nopagenumbers\footline={\ifnum\pageno>1\twelverm
\hss\folio\hss\fi}}
\def\leaderfill{\leaders\hbox to 1em{\hss.\hss}\hfill}
\def\ft#1#2{{\textstyle{{#1}\over{#2}}}}
\def\frac#1/#2{\leavevmode\kern.1em
\raise.5ex\hbox{\the\scriptfont0 #1}\kern-.1em/\kern-.15em
\lower.25ex\hbox{\the\scriptfont0 #2}}
\def\sfrac#1/#2{\leavevmode\kern.1em
\raise.5ex\hbox{\the\scriptscriptfont0 #1}\kern-.1em/\kern-.15em
\lower.25ex\hbox{\the\scriptscriptfont0 #2}}


\parindent=20pt
\def\narrow{\advance\leftskip by 40pt \advance\rightskip by 40pt}

\def\AB{\bigskip
        \centerline{\bf ABSTRACT}\medskip\narrow}
\def\nonarrower{\advance\leftskip by -40pt\advance\rightskip by -40pt}
\def\AE{\bigskip\nonarrower}

\def\boxit#1{\vbox{\hrule\hbox{\vrule\kern3pt
        \vbox{\kern3pt#1\kern3pt}\kern3pt\vrule}\hrule}}

\def\gtorder{\mathrel{\raise.3ex\hbox{$>$}\mkern-14mu
             \lower0.6ex\hbox{$\sim$}}}
\def\ltorder{\mathrel{\raise.3ex\hbox{$<$}|mkern-14mu
             \lower0.6ex\hbox{\sim$}}}
\def\dalemb#1#2{{\vbox{\hrule height .#2pt
        \hbox{\vrule width.#2pt height#1pt \kern#1pt
                \vrule width.#2pt}
        \hrule height.#2pt}}}

\font\fourteentt=cmtt10 scaled \magstep2
\font\fourteenbf=cmbx12 scaled \magstep1
\font\fourteenrm=cmr12 scaled \magstep1
\font\fourteeni=cmmi12 scaled \magstep1
\font\fourteenss=cmss12 scaled \magstep1
\font\fourteensy=cmsy10 scaled \magstep2
\font\fourteensl=cmsl12 scaled \magstep1
\font\fourteenex=cmex10 scaled \magstep2
\font\fourteenit=cmti12 scaled \magstep1
\font\twelvett=cmtt10 scaled \magstep1 \font\twelvebf=cmbx12
\font\twelverm=cmr12 \font\twelvei=cmmi12
\font\twelvess=cmss12 \font\twelvesy=cmsy10 scaled \magstep1
\font\twelvesl=cmsl12 \font\twelveex=cmex10 scaled \magstep1
\font\twelveit=cmti12
\font\tenss=cmss10
 
 \font\ninebf=cmbx7 scaled \magstep1
\font\ninerm=cmr7 scaled \magstep1 \font\ninei=cmmi7 scaled \magstep1
\font\ninesy=cmsy7 scaled \magstep1 
\font\eightrm=cmr7 scaled 1140 
 
\font\sevenbf=cmbx7 \font\sevenrm=cmr7 \font\seveni=cmmi7
\font\sevensy=cmsy7 

\catcode`@=11
\newskip\ttglue
\newfam\ssfam

\def\fourteenpoint{\def\rm{\fam0\fourteenrm}
\textfont0=\fourteenrm \scriptfont0=\tenrm \scriptscriptfont0=\sevenrm
\textfont1=\fourteeni \scriptfont1=\teni \scriptscriptfont1=\seveni
\textfont2=\fourteensy \scriptfont2=\tensy \scriptscriptfont2=\sevensy
\textfont3=\fourteenex \scriptfont3=\fourteenex \scriptscriptfont3=\fourteenex
\def\it{\fam\itfam\fourteenit} \textfont\itfam=\fourteenit
\def\sl{\fam\slfam\fourteensl} \textfont\slfam=\fourteensl
\def\bf{\fam\bffam\fourteenbf} \textfont\bffam=\fourteenbf
\scriptfont\bffam=\tenbf \scriptscriptfont\bffam=\sevenbf
\def\tt{\fam\ttfam\fourteentt} \textfont\ttfam=\fourteentt
\def\ss{\fam\ssfam\fourteenss} \textfont\ssfam=\fourteenss
\tt \ttglue=.5em plus .25em minus .15em
\normalbaselineskip=16pt
\abovedisplayskip=16pt plus 4pt minus 12pt
\belowdisplayskip=16pt plus 4pt minus 12pt
\abovedisplayshortskip=0pt plus 4pt
\belowdisplayshortskip=9pt plus 4pt minus 6pt
\parskip=5pt plus 1.5pt
\setbox\strutbox=\hbox{\vrule height12pt depth5pt width0pt}
\let\sc=\tenrm
\let\big=\fourteenbig \normalbaselines\rm}
\def\fourteenbig#1{{\hbox{$\left#1\vbox to12pt{}\right.\n@space$}}}

\def\twelvepoint{\def\rm{\fam0\twelverm}
\textfont0=\twelverm \scriptfont0=\ninerm \scriptscriptfont0=\sevenrm
\textfont1=\twelvei \scriptfont1=\ninei \scriptscriptfont1=\seveni
\textfont2=\twelvesy \scriptfont2=\ninesy \scriptscriptfont2=\sevensy
\textfont3=\twelveex \scriptfont3=\twelveex \scriptscriptfont3=\twelveex
\def\it{\fam\itfam\twelveit} \textfont\itfam=\twelveit
\def\sl{\fam\slfam\twelvesl} \textfont\slfam=\twelvesl
\def\bf{\fam\bffam\twelvebf} \textfont\bffam=\twelvebf
\scriptfont\bffam=\ninebf \scriptscriptfont\bffam=\sevenbf
\def\tt{\fam\ttfam\twelvett} \textfont\ttfam=\twelvett
\def\ss{\fam\ssfam\twelvess} \textfont\ssfam=\twelvess
\tt \ttglue=.5em plus .25em minus .15em
\normalbaselineskip=14pt
\abovedisplayskip=14pt plus 3pt minus 10pt
\belowdisplayskip=14pt plus 3pt minus 10pt
\abovedisplayshortskip=0pt plus 3pt
\belowdisplayshortskip=8pt plus 3pt minus 5pt
\parskip=3pt plus 1.5pt
\setbox\strutbox=\hbox{\vrule height10pt depth4pt width0pt}
\let\sc=\ninerm
\let\big=\twelvebig \normalbaselines\rm}
\def\twelvebig#1{{\hbox{$\left#1\vbox to10pt{}\right.\n@space$}}}

\def\tenpoint{\def\rm{\fam0\tenrm}
\textfont0=\tenrm \scriptfont0=\sevenrm \scriptscriptfont0=\fiverm
\textfont1=\teni \scriptfont1=\seveni \scriptscriptfont1=\fivei
\textfont2=\tensy \scriptfont2=\sevensy \scriptscriptfont2=\fivesy
\textfont3=\tenex \scriptfont3=\tenex \scriptscriptfont3=\tenex
\def\it{\fam\itfam\tenit} \textfont\itfam=\tenit
\def\sl{\fam\slfam\tensl} \textfont\slfam=\tensl
\def\bf{\fam\bffam\tenbf} \textfont\bffam=\tenbf
\scriptfont\bffam=\sevenbf \scriptscriptfont\bffam=\fivebf
\def\tt{\fam\ttfam\tentt} \textfont\ttfam=\tentt
\def\ss{\fam\ssfam\tenss} \textfont\ssfam=\tenss
\tt \ttglue=.5em plus .25em minus .15em
\normalbaselineskip=12pt
\abovedisplayskip=12pt plus 3pt minus 9pt
\belowdisplayskip=12pt plus 3pt minus 9pt
\abovedisplayshortskip=0pt plus 3pt
\belowdisplayshortskip=7pt plus 3pt minus 4pt
\parskip=0.0pt plus 1.0pt
\setbox\strutbox=\hbox{\vrule height8.5pt depth3.5pt width0pt}
\let\sc=\eightrm
\let\big=\tenbig \normalbaselines\rm}
\def\tenbig#1{{\hbox{$\left#1\vbox to8.5pt{}\right.\n@space$}}}
\let\rawfootnote=\footnote \def\footnote#1#2{{\rm\parskip=0pt\rawfootnote{#1}
{#2\hfill\vrule height 0pt depth 6pt width 0pt}}}

\def\tenfoot{\tenpoint\hskip-\parindent\hskip-.1cm}

\overfullrule=0pt
\twelvepoint
\def\sbullet{\raise.2em\hbox{$\scriptscriptstyle\bullet$}}
\nofirstpagenotwelve
\hsize=16.5 truecm
\baselineskip 15pt

\def\ft#1#2{{\textstyle{{#1}\over{#2}}}}

\oneandahalfspace
\
\rightline{CTP TAMU-62/92}
\rightline{September 1992}
\vskip 2truecm
\centerline{\bf A Note on Representations of the $w_\infty$
Algebra\footnote{$^\dagger$}{\tenfoot Contribution to the Proceedings of the
XXI DGM Conference, Nankai University, Tianjin, China.}}
\vskip 1.5truecm
\centerline{C.N. Pope\footnote{$^*$}{\tenfoot Supported in part
by the U.S. Department of Energy, under
grant DE-FG05-91ER40633.} and X.J. Wang}
\vskip 1.5truecm
\centerline{\it Center
for Theoretical Physics,
Texas A\&M University,}
\centerline{\it College Station, TX 77843--4242, USA.}

\vskip 1.5truecm
\AB\singlespace
 We explicitly demonstrate that the unitary representations of the $w_\infty$
algebra and its truncations are just the unitary representations of the
Virasoro algebra.

\np
\AE\oneandahalfspace

     The representation theory of the Virasoro algebra plays a crucial
r\^ole in our understanding of strings and conformal field theories.  An
important feature of the Virasoro algebra is that it admits a central
extension; it is only for positive values of the central charge $c$ that
non-trivial unitary representations can occur.

     Similar questions about the existence of unitary representations arise for
the various extended conformal algebras, known generically as $W$ algebras.
These are generated by higher-spin conformal currents in addition to the spin-2
current, the energy-momentum tensor, which generates the Virasoro algebra.  For
example, the $W_N$ algebras have currents of spins $2,3,\ldots,N$, for each of
which there is a non-trivial central extension [1,2].  These algebras are
non-linear, but there is an $N\to\infty$ limit, known as $W_\infty$, in which
linearity is regained [3]. However the $W_\infty$ algebra still retains the
other feature of the $W_N$ algebras mentioned above, namely that it has
non-trivial central extensions in all the higher-spin sectors.  There is a
classical limit of $W_\infty$, known as $w_\infty$ [4], in which all the
central extensions in the higher-spin sectors are lost, leaving only the usual
one in the Virasoro subalgebra.  The $w_\infty$ algebra can be truncated to
what one may call $w_N$, by setting all generators with spins greater than $N$
to zero. Among the extended conformal algebras, the $w_N$ and $w_\infty$
algebras are the simplest in structure.

      One might expect that the absence of central terms for higher spins in
the $w_N$ and $w_\infty$ algebras should result in their unitary
representations
being equivalent to the usual Virasoro unitary representations.  Since a
simple and explicit proof of this can be given, it seems worthwhile to
present it here.

    We shall consider the $w_N$ algebra first. It has conformal currents of
spins $2,3,\ldots ,N$. Let $w^i_m$ denote the $m$'th Fourier mode of the spin
$i+2$ current, where $i=0,1,2,\ldots,N-2$. They satisfy
$$
[w^i_m,\ w^j_n]=\big[ (j+1)m-(i+1)n\big]w^{i+j}_{m+n}+
{c\over{12}}(m^3-m)\delta^{i,0}
\delta^{j,0}\delta_{m+n,0}\ ,\eqno(1)
$$
where on the right-hand side $w^{i+j}_{m+n}$ is defined to be zero if
$i+j>N-2$. Note that the central extension is non-trivial only for the Virasoro
sector.

   The Hilbert space of physical states consists of highest-weight states and
their descendants. For the $w_N$ algebra, we can denote a highest-weight
state by $|\vec h\rangle$, where $\vec h=(h_0,h_1,h_2,\ldots,h_{N-2})$,
satisfying
$$
\eqalign{
w^i_0 |\vec h\rangle&=h_i|\vec h\rangle \ ,\cr
w^i_m |\vec h\rangle&=0,\qquad m\ge 1\ .\cr}\eqno(2)
$$
Descendant states are those created by acting on $|\vec h\rangle$ with a string
of $w^i_{-m}$'s, with $m\ge 1$. Unitarity of the Hilbert space
requires the matrix of inner products of any set of physical states to be
non-negative.  A particular case is the matrix of inner products of
states $w^i_{-m} |\vec h\rangle$, with $m\ge 1$ and $i=0,1,2,\ldots,N-2$;
$$
M_{ij}=\langle\vec h|w^i_mw^j_{-m}|\vec h\rangle\ .\eqno(3)
$$
To see what we can learn about unitarity from this matrix,  we look at the
$w_3$
and $w_4$ algebras first, which are the first two $w_N$ extensions of the
Virasoro algebra.

    The $w_3$ algebra takes the form
$$\eqalign{
[L_m,L_n]&=(m-n)L_{m+n}+{c\over {12}}(m^3-m)\delta_{m+n,0}\ ,\cr
[L_m,W_n]&=(2m-n)W_{m+n}\ ,\cr
[W_m,W_n]&=0\ ,\cr}\eqno(4)
$$
where $L_m$ and $W_m$ are the spin 2 and 3 generators. The matrix $M_{ij}$ is
given by
$$
(M_{ij})=\pmatrix{2mh_0+{\ft{c}{12}}(m^3-m) &3mh_1\cr
                        3mh_1 &0\cr}
\ .\eqno(5)
$$
One can easily see that for this matrix to be non-negative, it must be that
$h_1=0$.  Since there is no central term for $W_m$, and, as we have just
seen, the weight under $W_0$ is zero, it follows that all descendant states
involving $W_{-m}$'s are spurious.  The remaining descendant states are
those of the Virasoro Verma module, and the conditions for their unitarity
are standard, with no modification due to the higher-spin generators.
Thus the unitary representations of $w_3$ are the same as the unitary
Virasoro representations.

     The $w_4$ algebra takes the form
$$\eqalign{
[L_m,L_n]&=(m-n)L_{m+n}+{c\over {12}}(m^3-m)\delta_{m+n,0}\ ,\cr
[L_m,W_n]&=(2m-n)W_{m+n}\ ,\cr
[W_m,W_n]&=2(m-n)V_{m+n}\ ,\cr
[L_m,V_n]&=(3m-n)V_{m+n}\ ,\cr
[W_m,V_n]&=0=[V_m,V_n]\ ,\cr}\eqno(6)
$$
where $L_m$, $W_m$ and $V_m$ are the spin 2, 3 and 4 generators.
The corresponding matrix $M_{ij}$ is
$$
(M_{ij})=\pmatrix{ 2mh_0+{\ft{c}{12}}(m^3-m) &3mh_1 &4mh_2\cr
                        3mh_1 &4mh_2 &0\cr
                        4mh_2 &0 &0\cr}\ .
\eqno(7)
$$
For the outer $2\times2$ submatrix to be non-negative, we must have $h_2=0$.
Now consider the upper diagonal $2\times 2$ block.  Having found that $h_2=0$,
the requirement of non-negativity implies that $h_1=0$. So we have
$$
h_1=h_2=0\ .\eqno(9)
$$
Thus unitary $w_4$ representations are the same as unitary Virasoro
representations.

     The pattern that we have seen in the two examples above continues for
all the finite-$N$ $w_N$ algebras.  The matrix $M$ of inner products (3) for
the $w_N$ algebra is
$$
(M_{ij})=\pmatrix{
 2mh_0+{\ft{c}{12}}(m^3-m) &3mh_1 &4mh_2 &\cdots &Nmh_{N-2}\cr
                        3mh_1 &4mh_2 &5mh_3 &\cdots &0\cr
                        4mh_2 &5mh_3 &6mh_4 &\cdots &0\cr
                        \vdots&\vdots &\vdots &\ddots &\vdots\cr
                        N m h_{N-2} &0 &0 &\cdots &0\cr}
\ .\eqno(10)
$$
Following the same method as we used above, we get, for
the matrix to be non-negative, that
$$
h_1=h_2=\cdots =h_{N-3}=h_{N-2}=0\ ,\eqno(11)
$$
showing that the unitary representations of $w_N$ are the same as those of
the Virasoro algebra.

    The $w_\infty$ algebra is somewhat different from the finite-$N$ $w_N$
algebras, since it has infinitely-many conformal currents. The algebra takes
the same form as (1);
$$
[w^i_m,\ w^j_n]=\big[ (j+1)m-(i+1)n\big]w^{i+j}_{m+n}
+{c\over{12}}(m^3-m)\delta^{i,0}
\delta^{j,0}\delta_{m+n,0}\ ,\eqno(12)
$$
except that now we have $i=0,1,2,\ldots,\infty$.  A highest-weight state
$|\vec h\rangle$ satisfies
$$
\eqalign{
w^i_0 |\vec h\rangle&=h_i|\vec h\rangle \ ,\cr
w^i_m |\vec h\rangle&=0,\qquad m\ge 1\ .\cr}\eqno(13)
$$
The $M$ matrix is now infinite dimensional, given by
$$
(M_{ij})=\pmatrix{
 2mh_0+{\ft{c}{12}}(m^3-m) &3mh_1 &4mh_2 &\cdots \cr
                        3mh_1 &4mh_2 &5mh_3 &\cdots \cr
                        4mh_2 &5mh_3 &6mh_4 &\cdots \cr
                        \vdots&\vdots &\vdots  &\ddots\cr}
\ .\eqno(14)
$$
The method we used for the $w_N$ algebras, starting by considering the
outer $2\times 2$ submatrix, is not appropriate in this infinite-dimensional
case.   Instead, we first consider another subspace of physical states,
consisting of
$$
|\alpha\rangle=w^{2i}_{-2m}|\vec h\rangle\ ,\qquad
|\beta\rangle=(w^i_{-m})^2|\vec h\rangle\ ,\quad i\ge 1,\ m\ge 1 \ .\eqno(15)
$$
The inner products in this two-dimensional subspace are
$$\eqalign{
\langle\alpha|\alpha\rangle
&=\langle \vec h|w^{2i}_{2m} w^{2i}_{-2m}|\vec h\rangle
=4m(2i+1)h_{4i}\ ,\cr
\langle\beta|\beta\rangle
&=\langle \vec h|(w^{i}_{m})^2 (w^{i}_{-m})^2|\vec h\rangle
=8m^2(i+1)^2h_{2i}^2+4m^3(i+1)(2i+1)^2h_{4i}\ ,\cr
\langle\beta|\alpha\rangle
&=\langle \vec h|(w^{i}_{m})^2 w^{2i}_{-2m}|\vec h\rangle
=2m^2(4i+3)(2i+1)h_{4i}\ ,\cr}\eqno(16)
$$
where we have used the conditions (13) for highest-weight states.
The matrix of inner products in this subspace is
$$
(N_{ij})=\pmatrix{ \langle\alpha|\alpha\rangle &\langle\alpha|\beta\rangle\cr
\langle\beta|\alpha\rangle &\langle\beta|\beta\rangle\cr}\ .
\eqno(17)
$$
The unitarity of this space requires Det$(N_{ij})\ge 0$, {\it i.e.}
$$
\langle\alpha|\alpha\rangle \langle\beta|\beta\rangle -
\langle\alpha|\beta\rangle \langle\beta|\alpha\rangle\ge 0 \ .
\eqno(18)
$$
Substituting (16), we find
$$
{\rm Det} (N_{ij})=32m^3(2i+1)(i+1)^2h_{4i}h_{2i}^2-
4m^4(2i+1)^2(8i^2+12 i+5) h_{4i}^2\ .\eqno(19)
$$
Since $i>0$, it follows that for sufficiently large $m$, Det$(N_{ij})$ is
negative, unless $h_{4i}= 0$ when $i\ge 1$. In the above calculations, we see
that it is crucial that there are no central extensions in the $i\ge 1$
sectors.

     Having established that $h_{4i}=0$ for $i\ge1$, we may now apply the
previous argument that we used for the $w_N$ algebras, where we choose
$2\times2$ submatrices of the inner products $(M_{ij})$ given in (14), with
$h_{4i}$ as a diagonal entry.  This immediately shows that all the $h_i$'s
except $h_0$ are zero. For example, knowing $h_4=0$, the submatrix
$$
\pmatrix{4mh_2 &5mh_3\cr 5mh_3 &6mh_4\cr}\eqno(20)
$$
of (14) tells us that $h_3=0$.  The submatrix
$$
\pmatrix{
2mh_0+{\ft {c}{12}}(m^3-m) &4mh_2\cr 4mh_2 &6mh_4\cr}\eqno(21)
$$
then tells us that $h_2=0$, and so on.  Thus we arrive at the conclusion that
unitary representations of the $w_\infty$ algebra coincide with
Virasoro unitary representations.

     There are other interesting examples in the $w$ algebra family, namely the
$w_{1+N}$ algebras (including the $N\to\infty$ limit). They have the spin-1
current $w^{-1}$ besides all the currents in the $w_N$ algebra.  They take the
same form as (1), but with $i$ and $j$ now allowed to range from $-1$ to
$N-2$. They allow a central extension also in the spin-1 sector, with a central
charge $\tilde c$ that is independent of the spin-2 central charge $c$:
$$
[w^{-1}_m,\ w^{-1}_n]={\tilde c} m\delta_{m+n,0}\ .\eqno(22)
$$

      Denoting by $h_{-1}$ the $w^{-1}_0$ weight of a highest-weight state, the
$M$ matrix (3) is now given by
$$
(M_{ij})=\pmatrix{{\tilde c} m& mh_{-1} &2m h_0 &\cdots\cr
 mh_{-1} &2mh_0+{\ft{c}{12}}(m^3-m) &3mh_1 &\cdots \cr
                 2mh_0 &3mh_1 &4mh_2 &\cdots \cr
                 \vdots&\vdots &\vdots  &\ddots\cr}
\ .\eqno(23)
$$
The arguments that we used previously for $w_N$ and $w_\infty$ can be applied
{\it mutatis mutandis} to $w_{1+N}$ and $w_{1+\infty}$.  One immediately
concludes that here too, $0=h_1=h_2=h_3=\cdots$.  The new feature here is
that one must also demand the non-negativity of $2\times2$ submatrices in (23)
that have ${\tilde c} m$ as the upper-left entry.  By taking the case where
the lower-right entry is $4m h_2$, one can see that $h_0$ must be zero.  Then,
taking the upper-left $2\times2$ submatrix in (23), with $m=1$, one sees that
$h_{-1}$ must also vanish.  Thus we have
$$
0=h_{-1}=h_0=h_1=h_2=\cdots\ .\eqno(24)
$$
So for any $w_{1+N}$ algebra, including $w_{1+\infty}$, its unitary
representations are totally trivial.  The only state in its
Hilbert space is the $SL(2,R)$-invariant vacuum $|0\rangle$.

\bigskip
\centerline{\bf References}
\medskip
\item{[1]} A.B. Zamolodchikov, {\sl Teo. Mat. Fiz.} {\bf 65} (1985) 347.
\item{[2]} V.A. Fateev and S. Lukyanov, {\sl Int. J. Mod. Phys.} {\bf A3}
(1988) 507.
\item{[3]} C.N. Pope, L.J. Romans and X. Shen, {\sl Phys. Lett.} {\bf 236B}
(1990) 173; \nl {\sl Nucl. Phys.} {\bf B339} (1990) 191.
\item{[4]} I. Bakas, {\sl Phys. Lett.} {\bf 228B} (1989) 57.

\end